\documentclass[conference,9pt,letterpaper]{IEEEtran}

\usepackage{microtype}
\usepackage{amsmath,amssymb}
\usepackage{bbold}

\usepackage{textcomp}
\usepackage{newunicodechar}
\newunicodechar{⁡}{}

\usepackage[dvipsnames]{xcolor}
\colorlet{pink}{red!40}
\colorlet{linkblue}{cyan!60}
\usepackage{colortbl}

\usepackage{graphicx}
\usepackage{adjustbox}

\definecolor{Size7}{HTML}{EAF4FB}
\definecolor{Size8}{HTML}{F2ECFA}
\definecolor{Size32}{HTML}{FFF4DE}
\definecolor{Size671}{HTML}{F0F1F2}

\definecolor{Best7}{HTML}{CBE5F5}
\definecolor{Best8}{HTML}{DDCEF2}
\definecolor{Best32}{HTML}{F7DFA8}

\newcommand{\sizeSeven}{\cellcolor{Size7}\textbf{7B}}
\newcommand{\sizeEight}{\cellcolor{Size8}\textbf{8B}}
\newcommand{\sizeThirtyTwo}{\cellcolor{Size32}\textbf{32B}}
\newcommand{\sizeSixSevenOne}{\cellcolor{Size671}\textbf{671B}}

\newcommand{\second}[1]{\underline{#1}}

\usepackage{booktabs}
\usepackage{multirow}
\usepackage{makecell}
\usepackage{threeparttable}
\usepackage{tabularx}
\usepackage{array}
\usepackage{siunitx}

\providecommand{\Description}[1]{}
\usepackage{capt-of}

\providecommand{\best}[1]{\textbf{#1}}
\providecommand{\second}[1]{\underline{#1}}

\usepackage{pifont}

\usepackage{algorithm}
\usepackage{algpseudocode}

\algrenewcommand\algorithmicrequire{\textbf{Input:}}
\algrenewcommand\algorithmicensure{\textbf{Output:}}

\usepackage{tcolorbox}

\usepackage{comment}
\usepackage{lipsum}

\usepackage{cite}
\usepackage{hyperref}
\hypersetup{
    breaklinks=true,
    colorlinks=false,
    hidelinks,
    pdfauthor={Xianjing Zhu, Wenqi Lou, Yuanbo Wen, Teng Wang, Lei Gong, Qi Guo, Chao Wang, Xuehai Zhou},
    pdftitle={Rule-Guided and Trajectory-Aware Repair for LLM-Based HLS Generation},
    linkcolor=linkblue,
    citecolor=linkblue,
    urlcolor=black
}

\ExplSyntaxOn
\NewDocumentCommand \tempfnlabel { m }
{
    \int_zero_new:c { g_tmpfncounter_#1_int }
    \int_gset:cn { g_tmpfncounter_#1_int } { \value{footnote } }
}

\NewDocumentCommand \tempfncounter { m }
{
    \int_use:c { g_tmpfncounter_#1_int }
}
\ExplSyntaxOff

\newcommand{\ignore}[1]{}
\AtBeginDocument{%
  \providecommand\BibTeX{{%
    \normalfont B\kern-0.5em{\scshape i\kern-0.25em b}\kern-0.8em\TeX}}}

\begin{document}

\title{Rule-Guided and Trajectory-Aware Repair for LLM-Based HLS Generation}

\author{
  \IEEEauthorblockN{\normalsize
    Xianjing Zhu\textsuperscript{1}, Wenqi Lou\textsuperscript{1,2},
    Yuanbo Wen\textsuperscript{3}, Teng Wang\textsuperscript{1,2},
    Lei Gong\textsuperscript{1}, Qi Guo\textsuperscript{3},
    Chao Wang\textsuperscript{1}, Xuehai Zhou\textsuperscript{1}}
  \IEEEauthorblockA{\textsuperscript{1}University of Science and Technology of China\\
    \textsuperscript{2}Suzhou Institute for Advanced Research, University of Science and Technology of China\\
    \textsuperscript{3}SKL of Processors, Institute of Computing Technology, CAS, Beijing, China}
}
\maketitle
\vspace{-0.8cm}
\begin{abstract}

Large language models (LLMs) offer a promising path for directly generating functionally correct and synthesizable HLS-C from natural-language specifications.
However, existing approaches often struggle to turn multi-round tool feedback and reusable HLS knowledge into effective revisions, limiting end-to-end generation success.
We propose \textbf{MementoHLS}\footnote{
Code is available at: \url{https://github.com/ZD-zhu/MementoHLS}.}
an agentic HLS generation framework built around executable and falsifiable dual-memory-guided repair.
MementoHLS combines an \textbf{Executable Repair Contract Library (ERCL)}, which converts cross-task HLS knowledge into evidence-gated repair actions, with a \textbf{Reviewer-Grounded Falsification Ledger (RFL)}, which converts task-local repair history into trajectory evidence to suppress ineffective routes and preserve validated progress.
To ensure reliable outcome attribution, a Same-Candidate Reviewer binds Parse, Compile, Testbench (TB), and Synthesis (Synth) outcomes to the corresponding candidate, while a Single-Commit Arbiter commits at most one candidate per round.
Across Llama-3-8B and Qwen2.5-Coder-7B/32B, MementoHLS improves same-candidate TB$\land$Synth pass@1/pass@5 over Zero-shot by an average of 16.1/12.1 percentage points on HLS-Eval (94 tasks) and 15.1/10.4 on Bench4HLS (170 tasks).
The corresponding Bench4HLS gains average 11.0/8.5 percentage points across four strong proprietary LLMs.

\end{abstract}
%
\section{Introduction}

High-level synthesis (HLS) abstracts hardware design from RTL to algorithmic descriptions~\cite{Chen2024Allo,Ye2024HIDA}. Large language models (LLMs) support C/C++-to-HLS transformation~\cite{Collini2025C2HLSC,Xu2026HLSRewriter}, quality-of-results (QoR) optimization~\cite{Xiong2024HLSPilot,Li2026ChatHLS,Prakriya2025LIFT}, and direct HLS-C generation from natural-language specifications~\cite{gai2025exploring,Khan2025SAGEHLS,Zou2026HLSSeek}. Direct generation must satisfy behavioral, interface, and synthesis requirements while constructing the algorithm. We target candidates passing both the provided testbench (TB) and configured synthesis (Synth) flow under fixed task interfaces and validation assets, without exposing the reference implementation to the model. Same-candidate TB$\land$Synth success defines our operational success criterion and provides a starting point for subsequent QoR optimization.

\begin{figure}[t]
\centering
\includegraphics[width=\linewidth]{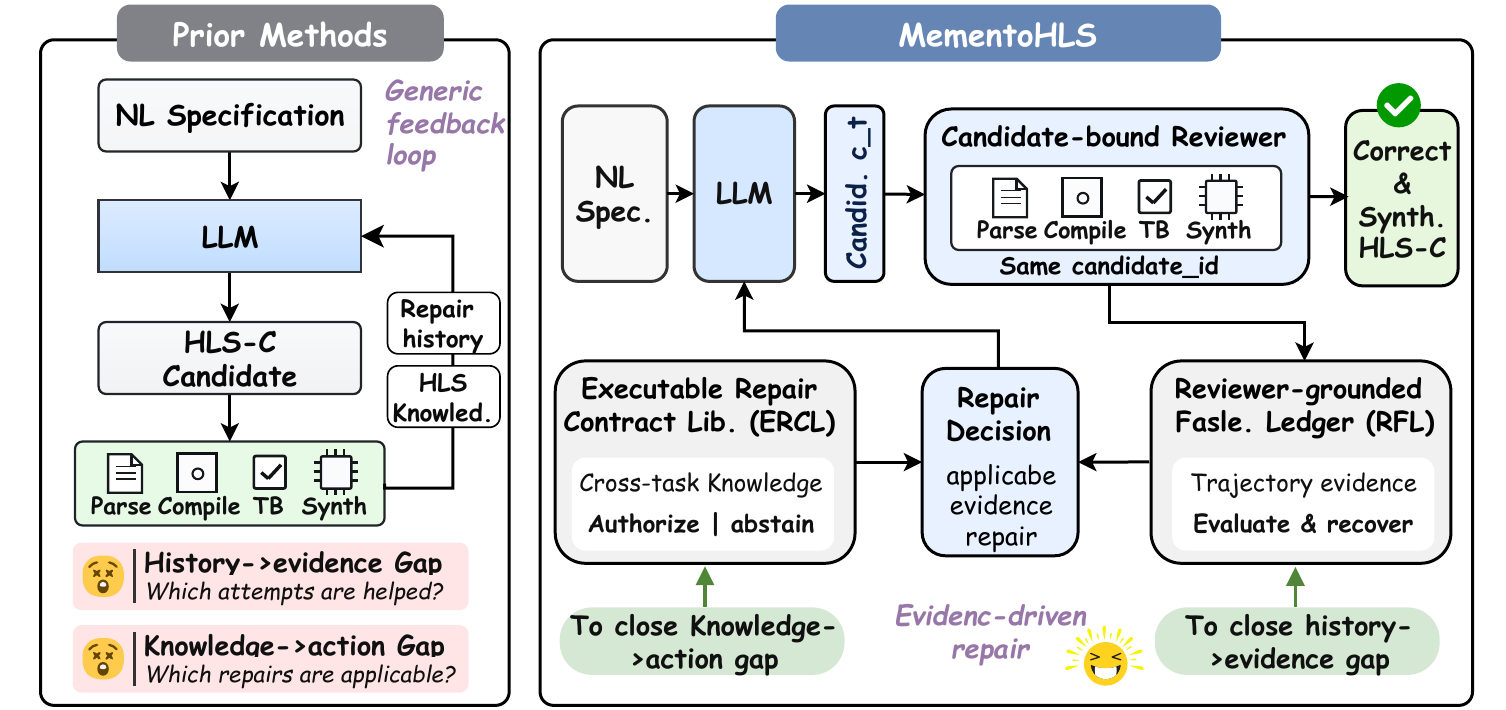}
\caption{Tool-feedback repair (left) and MementoHLS (right).}
\label{fig:introdution}
\end{figure}

Tool-guided iteration offers a way to address failures detected by these complementary checks~\cite{Chen2026ReflexHLS}. RTLFixer~\cite{tsai2024rtlfixer} combines retrieved guidance with compiler feedback; MEIC~\cite{Xu2024MEIC} performs iterative RTL debugging; HLS workflows use repair knowledge~\cite{Xu2026HLSRewriter} and structured design contracts~\cite{zhang2026contracthil}. Fig.~\ref{fig:introdution} (left) illustrates feedback-driven repair. Yet tool feedback alone does not determine which reusable repairs apply to a candidate or whether earlier edits improved it, leaving a \emph{knowledge-to-action gap} in repair applicability and a \emph{history-to-evidence gap} in assessing repair effectiveness.

\textbf{First, reusable HLS knowledge does not directly translate into reliable repair actions.} HLS-specific knowledge may describe synthesis restrictions, interface requirements, or numerical issues~\cite{Xu2025HLSTester}, but its applicability depends on the current candidate, task interface, data types, dimensions, tool feedback, and algorithmic constraints. Such a context-dependent applicability issue has also been observed in retrieval-augmented reasoning for hardware design specifications~\cite{Xing2026ChipMind}. A reliable repair mechanism must explicitly encode applicability evidence, prohibited conditions, permissible edits, design invariants, and validation obligations. To address this challenge, we introduce the \textbf{Executable Repair Contract Library (ERCL)}, which converts reusable cross-task HLS knowledge into evidence-gated and executable repair contracts.

\textbf{Second, raw interaction history does not provide reliable evidence about whether previous generation attempts have actually improved the design.}
 Across multiple rounds, superficial changes in logs may correspond to the same failure, while apparently different repairs may regress previously satisfied validation stages or revisit ineffective routes. Simply storing previous tool logs or model interactions cannot distinguish genuine progress from repeated failures, regression, or oscillation~\cite{Shinn2023Reflexion}. Reliable iterative repair requires attributing each observed outcome to the repair route that produced it, normalizing unstable tool feedback, and using this evidence to suppress ineffective actions while preserving validated progress. To address this challenge, we introduce the \textbf{Reviewer-Grounded Falsification Ledger (RFL)}, which converts task-local generation history into normalized trajectory evidence and uses actual tool outcomes to assign repair credit, suppress ineffective routes, preserve validated progress, and trigger recovery under stagnation.

We integrate ERCL and RFL into \textbf{MementoHLS}, an agentic framework for reliable LLM-based HLS generation, illustrated in Fig.~\ref{fig:introdution} (right). ERCL determines which reusable repair knowledge is applicable and authorized for the current generation failure, whereas RFL determines how the observed outcomes of previous repair attempts should constrain subsequent generation. For this dual-memory mechanism to operate reliably, however, every tool outcome must be unambiguously attributable to the exact candidate and repair action that produced it. We therefore introduce a \textbf{Same-Candidate Reviewer} to bind Parse, Compile, Testbench, and Synthesis outcomes to a unique candidate, together with a \textbf{Single-Commit Arbiter} that commits at most one candidate per round. 
These mechanisms provide reliable attribution for ERCL and RFL, forming a closed loop where reusable knowledge guides repair actions and trajectory evidence shapes subsequent generation.

In summary, this work makes the following contributions:

\begin{itemize}

\item We propose \textbf{MementoHLS}, an agentic framework for reliable LLM-based HLS generation that closes the loop among generation, tool feedback, and repair under a fixed LLM, while ensuring reliable outcome attribution through candidate-bound review and single-commit execution.

\item We introduce the \textbf{ERCL}, which transforms reusable cross-task HLS knowledge into evidence-gated, executable repair contracts with explicit applicability and validation constraints.

\item We introduce the \textbf{RFL}, which converts task-local multi-round repair history into tool-grounded trajectory evidence for repair credit assignment and adaptive trajectory control.


\item We evaluate MementoHLS on HLS-Eval \cite{AbiKaram2025HLSEval} and Bench4HLS \cite{Khan2026Bench4HLS}, demonstrating consistent improvements in same-candidate TB$\land$Synth success over the corresponding backbone
baselines across different model families and scales.

\end{itemize}

\section{Related Work} %

\subsection{Direct NL-to-HLS Generation}
\label{sec:direct-nl-to-hls}

Direct NL-to-HLS generation produces HLS C/C++ implementations from natural-language specifications without requiring existing source code.
Gai et al.~\cite{gai2025exploring} investigate fine-tuning, prompting, and feedback, while HLS-Eval~\cite{AbiKaram2025HLSEval} benchmarks compilation, functional correctness, and synthesizability.
SAGE-HLS~\cite{Khan2025SAGEHLS} employs synthesizable HLS data and serialized ASTs for structure-aware fine-tuning, while HLS-Seek~\cite{Zou2026HLSSeek} leverages functional and QoR signals through proxy-reward-based post-training and selective synthesis feedback.
ContractHIL-HLS~\cite{zhang2026contracthil} integrates structured design contracts with hardware-in-the-loop feedback for HLS generation and validation.

While these studies advance HLS generation through model adaptation or feedback-driven refinement, MementoHLS focuses on inference-time repair, combining executable repair contracts with trajectory evidence to guide multi-round revisions.

\subsection{C-to-HLS Transformation and Optimization}
\label{sec:c-to-hls-transformation}
A complementary line of work starts from existing C/C++ or HLS implementations and focuses on compatibility transformation, repair, and hardware optimization.
C2HLSC~\cite{Collini2025C2HLSC} transforms conventional C programs into HLS-compatible code through hierarchical decomposition and tool feedback, while HLSRewriter~\cite{Xu2026HLSRewriter} combines repair knowledge retrieval, code refactoring, and directive optimization.
RALAD~\cite{Xu2024RALAD} retrieves HLS examples and optimization knowledge, whereas AgRefactor~\cite{Zou2026AgRefactor} integrates specialized agents, automated refactoring, and cross-task memory.
For QoR optimization, HLSPilot~\cite{Xiong2024HLSPilot} combines profiling with design-space exploration, while LIFT~\cite{Prakriya2025LIFT} focuses on structure-aware pragma generation.

These approaches benefit from existing implementations that provide algorithmic and structural guidance~\cite{Li2026ChatHLS,europar26-two}.
In contrast, MementoHLS starts from natural-language specifications and employs multi-round repair to obtain functionally correct and synthesizable HLS implementations, providing a foundation for subsequent QoR optimization.

\subsection{Tool-Guided EDA Agents and Iterative Repair}
\label{sec:tool-guided-repair}

Recent LLM-based EDA agents have increasingly integrated tool execution and feedback into automated design workflows~\cite{Blocklove2025AutoChip,Ho2025VerilogCoder,Zhao2025MAGE}.
ChatEDA~\cite{wu2024chateda} coordinates EDA tools through LLM-based planning and execution.
RTLFixer~\cite{tsai2024rtlfixer} combines retrieval-augmented guidance with iterative tool feedback to repair RTL syntax errors, while MEIC~\cite{Xu2024MEIC} employs iterative LLM-based debugging for RTL repair.
In the HLS domain, ChatHLS~\cite{Li2026ChatHLS} integrates multi-agent generation, tool-guided debugging, and QoR optimization.

Building on tool-guided repair, MementoHLS couples evidence-gated executable repair contracts with task-local trajectory evidence to guide repair decisions.
Its candidate-bound Reviewer and single-commit execution further associate each repair attempt with its corresponding Parse, Compile, TB, and Synthesis outcomes, enabling ineffective repair routes to be identified and suppressed.

\begin{figure*}
    \centering
    \includegraphics[width=1.0\linewidth]{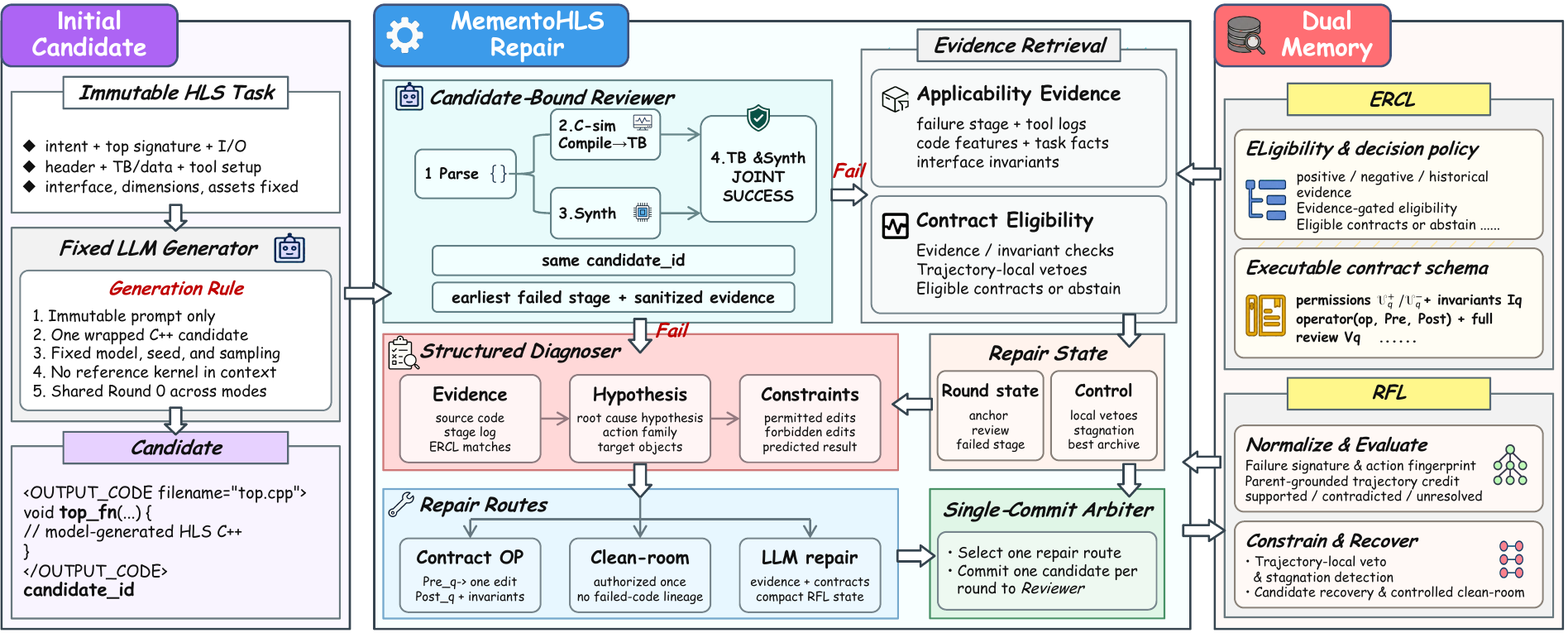}
    \vspace{-0.6cm}
    \caption{Overview of MementoHLS: dual-memory-guided repair with candidate-bound validation.}
    \label{fig:framework}
    \vspace{-4pt}
\end{figure*}

\section{MementoHLS Framework}\label{sec_framework}
\subsection{MementoHLS Overview}

This section presents \textsc{MementoHLS}, an evidence-driven repair
framework deployed around a fixed LLM. MementoHLS integrates reusable repair knowledge from ERCL and task-local trajectory evidence from RFL to guide iterative HLS candidate repair.

We represent an HLS task as $x=(d,I,\mathcal{A},\delta)$,
%
where $d$ denotes the natural-language design intent, $I$ denotes immutable
interface artifacts, $\mathcal{A}$ denotes validation assets, and $\delta$
specifies the tool configuration. 
Given a dataset $\mathcal{D}$, the projection $\pi_{\mathcal{D}}(x)$ determines the content visible to the model. In contrast, the $\mathtt{Reviewer}$ retains the complete task $x$, while the reference implementation never enters the model context.

A repair trajectory consists of uniquely identified candidates:
\begin{equation}
c_t=(\mathrm{id}_t,z_t,\mathrm{code}_t,\mathrm{parent}_t,\mathrm{origin}_t)
\label{eq:candidate-definition}
\end{equation}
where \(\mathrm{id}_t\) denotes the unique candidate identifier and \(z_t\) the full generated payload. Upon successful parsing, \(\mathrm{code}_t\) is the extracted source code; otherwise, \(\mathrm{code}_t=\bot\). $\mathrm{parent}_t$ identifies the repair source for credit comparison
and $\mathrm{origin}_t$ records the generation route. The complete artifact is
stored in trajectory history $\mathcal{H}$, while RFL maintains compressed
trajectory states.

As shown in Fig.~\ref{fig:framework}, given an NL-to-HLS task, MementoHLS first generates an initial candidate using
the fixed LLM. In each subsequent round, the $\mathtt{Reviewer}$ evaluates the committed
candidate through Parse, Compile, TB, and Synth, and identifies the earliest
valid failure. Based on candidate-bound evidence, ERCL retrieves applicable
repair knowledge, while RFL provides trajectory evidence including previous
outcomes, local vetoes, and candidate states. The $\mathtt{Diagnoser}$ generates a
structured repair hypothesis, and the Single-Commit $\mathtt{Arbiter}$ selects one repair
route to produce the next candidate. The new candidate is evaluated by the
same tool chain, and the outcome is returned to RFL for trajectory-level credit
assignment.
In this process, \textbf{ERCL governs what repair can be attempted, while RFL
governs how previous repair outcomes should influence subsequent generation.}
Candidate-bound validation and single-commit execution provide reliable
outcome attribution for both mechanisms.

\subsection{Candidate-Bound Review and Repair Execution}
\label{sec:reviewer}

ERCL and RFL rely on reliable evidence attribution. Therefore, MementoHLS uses
candidate-bound review and single-commit execution as the execution substrate
for closed-loop repair.
For each candidate $c_t$, the $\mathtt{Reviewer}$ returns a complete review
record:
\begin{equation}
\mathcal{R}(c_t)=
(\mathbf{v}_t,\mathbf{e}_t),
\qquad
\mathbf{v}_t=(P_t,C_t,T_t,S_t)
\label{eq:review-record}
\end{equation}
where the four states indicate whether the candidate passes Parse ($P_t$), Compile ($C_t$),
TB ($T_t$), and Synth ($S_t$). The evidence $\mathbf{e}_t$ is bound to the same
\texttt{candidate\_id} and contains tool outcomes and diagnostic information.
The $\mathtt{Reviewer}$ identifies the first failing stage in the standard HLS kernel development flow,  i.e., \(\mathrm{Parse}\rightarrow\mathrm{Compile}\rightarrow\mathrm{TB}\rightarrow\mathrm{Synth}\). 
This ordering defines the repair target and does not terminate tool execution. 
We define same-candidate joint success (TB$\land$Synth) as \(J_t=T_t\land S_t\), where \(T_t\) and \(S_t\) denote the TB and Synth outcomes of the same candidate \(c_t\), respectively.

For each repair round, the $\mathtt{Diagnoser}$ generates a structured hypothesis based
on the candidate, failure evidence, ERCL contracts, and RFL state. It produces
a root-cause hypothesis, action family, target objects, and permitted or
prohibited modifications. The diagnosis is treated only as a repair hypothesis
rather than a confirmed cause; repair effectiveness is determined solely by
subsequent tool outcomes.
The Single-Commit $\mathtt{Arbiter}$ selects at most one route among deterministic
contract execution, authorized clean-room generation, and local LLM repair.
This guarantees that each observed outcome can be uniquely attributed to the
repair route that produced the candidate.
Together, candidate-bound review and single-commit execution provide the
auditable evidence substrate required by ERCL and RFL.

\subsection{Executable Repair Contract Library}
\label{sec:ercl}

ERCL converts reusable cross-task HLS knowledge from passive textual guidance
into executable repair contracts. Instead of directly injecting retrieved
knowledge into the LLM prompt, ERCL represents repair knowledge with explicit
applicability boundaries, execution constraints, and validation obligations.

Each contract is represented as: 
\begin{equation}
q =
\left(
\mathrm{id}_q,
\phi_q,
\mathcal{E}_q,
\mathcal{U}_q,
\mathcal{I}_q,
\mathcal{O}_q,
\mathcal{V}_q,
\omega_q,
\kappa_q
\right)
\label{eq:repair-contract}
\end{equation}
where $\phi_q$ denotes the target failure stage;
$\mathcal{E}_q=(\mathcal{E}_q^{+},\mathcal{E}_q^{-},\mathcal{E}_q^{h})$
denotes positive, negative, and historical applicability evidence;
$\mathcal{U}_q=(\mathcal{U}_q^{+},\mathcal{U}_q^{-})$ defines permitted and
prohibited modifications; $\mathcal{I}_q$ specifies preserved design
invariants; $\mathcal{O}_q=(\mathrm{op}_q,\mathrm{Pre}_q,\mathrm{Post}_q)$
describes an optional deterministic operator and its preconditions and
postconditions; and $\mathcal{V}_q$ denotes
the required $\mathtt{Reviewer}$ validation obligation. Different from workflow memories
that reuse previous procedures from task trajectories
\cite{Wang2025AgentWorkflowMemory}, ERCL provides explicit repair authority
through applicability evidence, modification constraints, invariants, and
validation requirements.

At runtime, ERCL determines contract eligibility using tool feedback,
candidate code, task facts, and trajectory history. It filters contracts with
matched negative evidence, conflicting invariants, insufficient support, or
RFL local vetoes. Immutable interface constraints, precise diagnostics, and
concrete code features take precedence over textual similarity. Therefore, a
retrieved match does not itself grant repair authority. Only contracts with
sufficient evidence enter the eligible set $Q_t^{\mathrm{ret}}$; otherwise,
ERCL returns $\bot$ and abstains from contract-driven repair.

For an eligible contract, ERCL authorizes only the predefined transformation:
\begin{equation}
\mathrm{Pre}_q
\rightarrow
\mathrm{op}_q
\rightarrow
\mathrm{Post}_q 
\end{equation}
The $\mathtt{Executor}$ verifies $\mathrm{Pre}_q$, applies $\mathrm{op}_q$, and checks
$\mathrm{Post}_q$. Any failed precondition, execution exception, invariant
violation, or \texttt{no-change} result causes the contract to abstain,
preventing unconstrained LLM modifications.
After execution, the modified candidate must satisfy the complete validation
obligation:
\begin{equation}
\mathcal{V}_q =
\mathrm{Parse}
+
\mathrm{Compile}
+
\mathrm{TB}
+
\mathrm{Synth}
\end{equation}
The validation result is obtained through the same Candidate-Bound $\mathtt{Reviewer}$,
ensuring that contract effectiveness is determined by actual tool outcomes.
Therefore, ERCL follows the procedure:
\[
\mathrm{retrieve}
\rightarrow
\mathrm{authorize}
\rightarrow
\mathrm{execute}
\rightarrow
\mathrm{validate}
\]

\begin{figure}[t]
    \centering
        \includegraphics[width=1\linewidth]{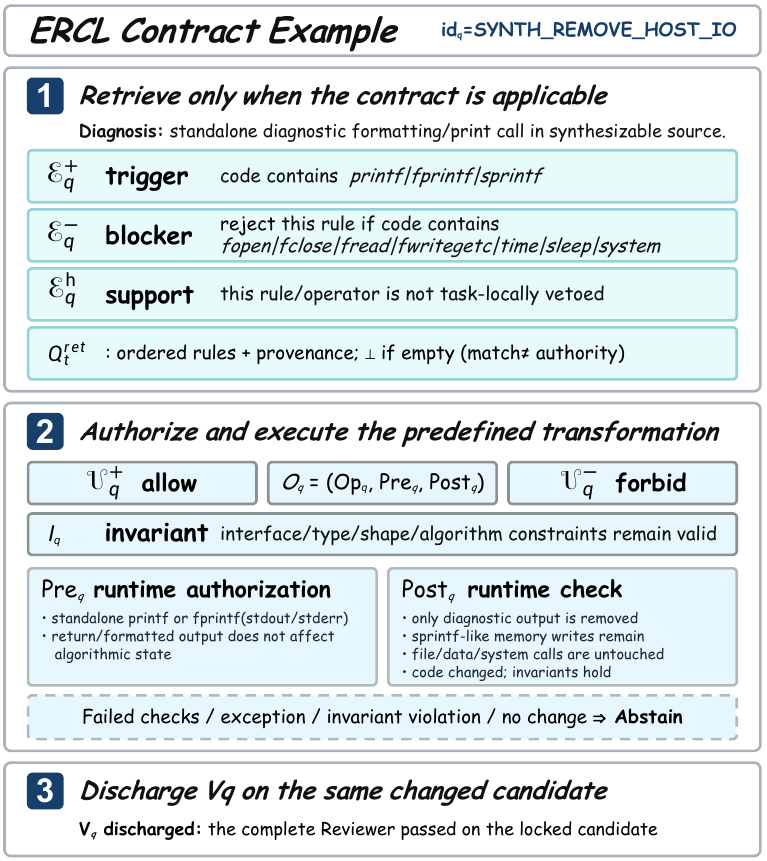}
    \vspace{-0.6cm}
    \caption{An example executable repair contract in ERCL.}
    \label{fig:ercl}
    \vspace{-0.4cm}
\end{figure}

Fig.~\ref{fig:ercl} illustrates this contract semantics using the
synthesis-stage contract \texttt{SYNTH\_REMOVE\_HOST\_IO}, showing how ERCL
combines applicability evidence, execution constraints, and validation
obligations to enable safe and auditable repair.

\subsection{Reviewer-Grounded Falsification Ledger and Closed-Loop Control}
\label{sec:rfl}

The Reviewer-Grounded Falsification Ledger (RFL) is a task-local trajectory
memory that converts multi-round repair history into normalized evidence for
guiding subsequent HLS generation and repair. Unlike raw tool logs or
interaction histories, RFL evaluates repair routes according to actual
$\mathtt{Reviewer}$ outcomes, assigns credit to executed actions, and constrains future
repair decisions. The RFL is isolated by
$(\mathrm{task},\mathrm{sample},\mathrm{seed},\mathrm{mode})$ and does not
update the global ERCL. Here, \emph{falsification} refers to locally testing
whether a repair hypothesis is supported by actual tool outcomes.

The RFL maintains repair trajectories through four stages:
\emph{normalize}, \emph{evaluate}, \emph{constrain}, and \emph{recover}.

\textbf{Normalize.}
For a failed candidate $c_t$, the RFL removes unstable fields from tool
feedback, including paths, source locations, timestamps, and run names, and
constructs the failure signature:
\begin{equation}
\sigma_t =
\operatorname{Hash}
\left(
f(c_t),
\operatorname{Norm}
(
\mathbf{e}_t[f(c_t)]
)
\right)
\label{eq:failure-signature}
\end{equation}
where $\mathbf{e}_t[f(c_t)]$ denotes the evidence of the earliest failed stage.
The RFL further constructs an action fingerprint from the normalized action
family, contract/operator identifier, and target object to identify repeated
failures and ineffective repair routes.

\textbf{Evaluate.}
The RFL compares the repaired candidate with its actual parent candidate and
assigns credit to the \emph{actual submitted repair route}, rather than the
natural-language diagnosis. A route is labeled \emph{supported} if it improves
the validation frontier without regression or invariant violation, and
\emph{contradicted} if it preserves the same failure, degrades previous
progress, or violates constraints; otherwise, it remains \emph{unresolved}.

\textbf{Constrain.}
When a repair contract, action family, or route is falsified with an
identifiable target, the RFL records it into a trajectory-local negative
ledger indexed by failure stage and failure signature. These local vetoes
constrain subsequent ERCL retrieval and LLM repair decisions, while the global
ERCL remains unchanged. The effective repair space is therefore jointly
determined by cross-task knowledge and trajectory evidence:
\begin{equation}
\mathcal{A}_t =
\operatorname{Eligible}_{ERCL}(c_t,\mathbf{e}_t)
-
\operatorname{Veto}_{RFL}(\mathcal{H}_t)
\label{eq:repair-space}
\end{equation}

\textbf{Recover.}
The RFL detects stagnation from repeated failure signatures, repeated repair
routes, A--B--A oscillation, conflicting diagnoses, or missing progress.
When stagnation occurs, RFL supports candidate recovery or controlled
clean-room generation. The recovered candidate follows the same single-commit
and Reviewer validation process, and its outcome is incorporated into the
trajectory evidence.

Together, ERCL and RFL form a closed-loop repair process: ERCL provides
reusable repair opportunities according to current candidate evidence, while
RFL continuously updates their trajectory-local validity based on actual
execution outcomes. Thus, subsequent repair decisions are jointly constrained
by cross-task repair knowledge and task-local trajectory evidence.


\renewcommand{\arraystretch}{1.0}

\setlength{\extrarowheight}{0.3pt}

\setlength{\tabcolsep}{7.2pt}

\newlength{\TypeColWidth}
\newlength{\ModelColWidth}
\newlength{\SizeColWidth}

\setlength{\TypeColWidth}{3.2cm}
\setlength{\ModelColWidth}{3.0cm}
\setlength{\SizeColWidth}{1.5cm}

\begin{table*}[t]
\centering
\caption{HLS-Eval pass@1 (p@1) and pass@5 (p@5) success rates (\%) for parsing, compilation, testbench (TB), synthesis, and joint TB$\land$Synth. Color denotes size. Bold/underline mark first/second per size and metric in our runs.}
\label{tab:hls_eval_results}

\begin{minipage}{\textwidth}
\centering

\resizebox{\textwidth}{!}{
\begin{tabular}{
>{\raggedright\arraybackslash}p{\TypeColWidth}
>{\raggedright\arraybackslash}p{\ModelColWidth}
>{\centering\arraybackslash}p{\SizeColWidth}
*{10}{c}
}
\toprule

\multirow{3}{*}{\textbf{Type}} &
\multirow{3}{*}{\textbf{Model}} &
\multirow{3}{*}{\textbf{Size}} &
\multicolumn{10}{c}{\textbf{HLS-Eval \cite{AbiKaram2025HLSEval} (\%)}} \\

\cmidrule(lr){4-13}

& & &
\multicolumn{2}{c}{\textbf{Parse}} &
\multicolumn{2}{c}{\textbf{Compile}} &
\multicolumn{2}{c}{\textbf{TB}} &
\multicolumn{2}{c}{\textbf{Synth}} &
\multicolumn{2}{c}{\textbf{TB$\land$Synth}} \\

\cmidrule(lr){4-5}
\cmidrule(lr){6-7}
\cmidrule(lr){8-9}
\cmidrule(lr){10-11}
\cmidrule(lr){12-13}

& & &
\textbf{p@1} & \textbf{p@5} &
\textbf{p@1} & \textbf{p@5} &
\textbf{p@1} & \textbf{p@5} &
\textbf{p@1} & \textbf{p@5} &
\textbf{p@1} & \textbf{p@5} \\

\midrule


\makecell[l]{Base Model}
& Llama-3
& \sizeEight
& \second{99.8} & \best{100.0}
& 64.3 & 80.9
& 45.3 & 59.6
& 59.1 & 76.6
& 41.7 & 57.4
\\

\midrule


\multirow{2}{*}{\makecell[l]{Code\\Models}}

& Qwen2.5-Coder
& \sizeThirtyTwo
& \second{99.1} & \best{100.0}
& \second{88.6} & \second{94.3}
& \second{59.1} & \second{65.9}
& \second{82.3} & \second{93.2}
& \second{56.4} & \second{64.8}
\\

& Qwen2.5-Coder
& \sizeSeven
& \second{99.1} & \best{100.0}
& 78.7 & 89.4
& 55.5 & 67.0
& 74.9 & 88.3
& 53.8 & 65.9
\\

\midrule


\multirow{2}{*}{\textbf{SAGE-HLS \cite{Khan2025SAGEHLS}}}

& \mbox{Llama-3 + FT$^{\dagger}$}
& \sizeEight
& 95.3 & \best{100.0}
& 60.6 & 80.9
& 46.6 & 61.7
& 53.4 & 79.8
& 40.9 & 59.6
\\

& \mbox{Qwen2.5-Coder + FT$^{\dagger}$}
& \sizeSeven
& \best{100.0} & \best{100.0}
& 75.1 & \second{94.7}
& 53.6 & 69.1
& 67.2 & 93.6
& 46.8 & 67.0
\\

\midrule


\textbf{ContractHIL-HLS \cite{zhang2026contracthil}}
& DeepSeek V3
& \sizeSixSevenOne
& 96.8 & -
& 87.4 & -
& 70.4 & 76.6
& 94.7 & -
& - & -
\\

\midrule


\textbf{Gai et al. \cite{gai2025exploring}}
& CodeLlama
& \sizeSeven
& - & -
& 6.3 & 48.8
& 1.4 & 27.9
& - & -
& - & -
\\

\midrule


\multirow{2}{*}{\textbf{HLS-Seek \cite{Zou2026HLSSeek}}}

& \mbox{HLS-Seek (SFT)}
& \sizeSeven
& - & -
& 77.5 & 88.4
& 50.7 & 60.5
& - & -
& - & -
\\

& \mbox{HLS-Seek (GRPO)}
& \sizeSeven
& - & -
& 84.7 & 95.3
& 68.9 & 81.4
& - & -
& - & -
\\

\midrule


\multirow{5}{*}{\textbf{MementoHLS}}

& Qwen2.5-Coder
& \sizeSeven
& \best{100.0} & \best{100.0}
& \second{90.4} & 94.6
& \second{66.6} & \second{74.4}
& \best{89.7} & \second{96.8}
& \best{65.7} & \second{74.4}
\\

& \mbox{Qwen2.5-Coder + FT$^{\dagger}$}
& \sizeSeven
& \best{100.0} & \best{100.0}
& \best{92.5} & \best{97.8}
& \best{67.4} & \best{79.7}
& \second{86.8} & \best{97.9}
& \second{63.1} & \best{78.7}
\\

& Qwen2.5-Coder
& \sizeThirtyTwo
& \best{100.0} & \best{100.0}
& \best{96.8} & \best{98.9}
& \best{77.2} & \best{81.9}
& \best{94.4} & \best{98.9}
& \best{75.5} & \best{81.9}
\\

& Llama-3
& \sizeEight
& \best{100.0} & \best{100.0}
& \best{81.5} & \second{91.5}
& \best{59.4} & \best{69.1}
& \best{80.2} & \second{92.6}
& \best{58.9} & \best{68.1}
\\

& \mbox{Llama-3 + FT$^{\dagger}$}
& \sizeEight
& \best{100.0} & \best{100.0}
& \second{77.4} & \best{94.6}
& \second{55.5} & \second{68.0}
& \second{70.4} & \best{95.7}
& \second{50.6} & \second{65.9}
\\

\bottomrule
\end{tabular}
}


\vspace{1.5mm}

\parbox{\linewidth}{
\footnotesize
\raggedright

$^{\dagger}$FT denotes LoRA fine-tuning on the training corpus released by SAGE-HLS \cite{Khan2025SAGEHLS}.
``--'' denotes unreported settings, which we do not reproduce because the implementation is not publicly available.
}

\end{minipage}
\end{table*}

\begin{table}[t]
\centering
\caption{Bench4HLS success (pass@1/pass@5). Color denotes size. Bold/underline mark first/second per size and metric.}
\label{tab:bench4hls_results}

\scriptsize
\setlength{\tabcolsep}{2.2pt}
\renewcommand{\arraystretch}{1.05}

\resizebox{\columnwidth}{!}{
\begin{tabular}{l l c *{6}{c}}
\toprule

\multirow{3}{*}{\textbf{Type}} &
\multirow{3}{*}{\textbf{Model}} &
\multirow{3}{*}{\textbf{Size}} &
\multicolumn{6}{c}{\textbf{Bench4HLS~\cite{Khan2026Bench4HLS} (\%)}} \\

\cmidrule(lr){4-9}

& & &
\multicolumn{2}{c}{\textbf{TB}} &
\multicolumn{2}{c}{\textbf{Synth}} &
\multicolumn{2}{c}{\textbf{TB$\land$Synth}} \\

\cmidrule(lr){4-5}
\cmidrule(lr){6-7}
\cmidrule(lr){8-9}

& & &
\textbf{p@1} & \textbf{p@5} &
\textbf{p@1} & \textbf{p@5} &
\textbf{p@1} & \textbf{p@5} \\

\midrule


\makecell[l]{Base Model}
& Llama-3
& \sizeEight
& 15.8 & 32.9
& 58.9 & 95.3
& 15.5 & 32.9
\\

\midrule


\multirow{2}{*}{\makecell[l]{Code\\Models}}
& Qwen2.5-Coder
& \sizeThirtyTwo
& \second{49.8} & \second{60.0}
& \second{85.9} & \second{97.6}
& \second{48.6} & \second{59.4}
\\

& Qwen2.5-Coder
& \sizeSeven
& 26.7 & 51.1
& 52.1 & 91.8
& 26.4 & 50.6
\\

\midrule


\multirow{2}{*}{\textbf{SAGE-HLS~\cite{Khan2025SAGEHLS}}}
& \mbox{Llama-3 + FT$^{\dagger}$}
& \sizeEight
& 28.9 & 42.9
& \second{80.4} & 96.5
& 28.9 & 42.3
\\

& \mbox{Qwen2.5-Coder + FT$^{\dagger}$}
& \sizeSeven
& 32.2 & 51.1
& 80.3 & \best{98.8}
& 31.4 & 48.8
\\

\midrule


\multirow{5}{*}{\textbf{MementoHLS}}
& Qwen2.5-Coder
& \sizeSeven
& \second{43.6} & \best{58.2}
& \second{86.1} & \second{98.2}
& \second{42.3} & \best{57.0}
\\

& \mbox{Qwen2.5-Coder + FT$^{\dagger}$}
& \sizeSeven
& \best{45.1} & \second{54.7}
& \best{97.2} & \best{98.8}
& \best{44.1} & \second{52.9}
\\

& Qwen2.5-Coder
& \sizeThirtyTwo
& \best{61.7} & \best{68.2}
& \best{96.9} & \best{98.8}
& \best{61.0} & \best{66.4}
\\

& Llama-3
& \sizeEight
& \second{32.5} & \best{50.6}
& 79.2 & \second{98.2}
& \second{32.4} & \best{50.6}
\\

& \mbox{Llama-3 + FT$^{\dagger}$}
& \sizeEight
& \best{35.6} & \second{49.4}
& \best{93.7} & \best{98.8}
& \best{35.2} & \second{48.8}
\\

\bottomrule
\end{tabular}
}

\end{table}

\section{Experiments}




We evaluate MementoHLS on two NL-to-HLS benchmarks to investigate its effectiveness, component contributions, and repair efficiency. We also examine the latency of generated designs relative to hand-designed benchmark references. Specifically, we address three research questions:

\noindent\textbf{RQ1:} How effectively does MementoHLS improve end-to-end HLS generation across model families, parameter scales, and domain-adaptation settings, including proprietary LLMs?

\noindent\textbf{RQ2:} What are the individual and combined contributions of ERCL and RFL to HLS generation success?

\noindent\textbf{RQ3:}  How does generation success evolve over multiple repair rounds, and what are the associated costs in repair rounds, LLM calls, token usage, and HLS tool runtime?

\textit{Benchmarks.} We evaluate MementoHLS on HLS-Eval~\cite{AbiKaram2025HLSEval} and Bench4HLS~\cite{Khan2026Bench4HLS}. HLS-Eval contains 94 manually verified designs spanning diverse workloads and application domains, while Bench4HLS comprises 170 manually curated and validated designs emphasizing larger and more structurally complex workloads. Together, they provide complementary evaluation settings across different task scales and complexity regimes.

\textit{Models and Baselines.} We evaluate Llama-3-8B and Qwen2.5-Coder-7B/32B, along with LoRA-adapted variants trained on the SAGE-HLS~\cite{Khan2025SAGEHLS} corpus. Baselines include Zero-shot generation, feedback-only repair, ERCL/RFL ablations, and published results from representative NL-to-HLS methods. We additionally evaluate four proprietary LLMs.

\textit{Evaluation Protocol and Metrics.} Each task is evaluated using five independent trajectories with different random seeds. Zero-shot uses only the initial candidate (Round 0), while iterative configurations allow up to five repair rounds, stopping early upon joint TB$\land$Synth success. We report pass@1 as the success rate across individual trajectories and pass@5 as the fraction of tasks with at least one successful trajectory. In addition to stage-wise success, our primary metric is same-candidate TB$\land$Synth success, as defined in Sec.~\ref{sec_framework}.

\textit{Implementation.} MementoHLS is implemented in Python with vLLM on Ubuntu 22.04 and an NVIDIA A100 80 GB GPU. Open-weight models use an 8192-token context and a sampling temperature of 0.7 for generation and repair. The ERCL contains 71 contracts (6 Parse, 23 Compile, 27 TB, and 15 Synth) and is frozen before evaluation, without using benchmark failures, reference implementations, or evaluation outcomes. HLS validation uses Vitis HLS 2024.2 and Vivado targeting the Xilinx \texttt{xczu9eg-ffvb1156-2-e} device under a 5 ns clock constraint. Proprietary LLMs use matched within-model API settings, with native budgets different from those of open-weight runs.

\subsection{Generation Effectiveness}

To answer RQ1, we compare MementoHLS with the corresponding backbone baselines on HLS-Eval and Bench4HLS. 
As shown in Table~\ref{tab:hls_eval_results} and Table~\ref{tab:bench4hls_results}, MementoHLS consistently improves
same-candidate TB$\land$Synth pass@1 and pass@5 across both benchmarks and all
evaluated backbones, with gains of up to 19.1 and 17.7 percentage points,
respectively. These improvements persist from 7/8B models to the stronger 32B
backbone, indicating that MementoHLS improves end-to-end generation
reliability rather than merely compensating for limited model capability.


The results on domain-adapted backbones further demonstrate the
complementarity between MementoHLS and offline HLS adaptation.
While LoRA fine-tuning injects HLS-specific knowledge into model
parameters, MementoHLS leverages candidate-level Parse, Compile,
TB, and Synth feedback to perform inference-time repair.
Table~\ref{tab:hls_eval_results} lists the available published
stage-wise scores for Gai et~al., ContractHIL-HLS, and HLS-Seek.
For same-candidate TB$\land$Synth, we compare MementoHLS with its
corresponding backbone baselines evaluated under our protocol.

\begin{table}[!ht]
\centering
\caption{Effectiveness of MementoHLS on proprietary LLMs. Same-candidate TB$\land$Synth success rates (\%) on HLS-Eval and Bench4HLS. Each entry shows zero-shot $\rightarrow$ MementoHLS, with MementoHLS results in bold.}
\label{tab:closed_models}
\footnotesize
\setlength{\tabcolsep}{5pt}
\begin{tabular}{@{}lccc@{}}
\toprule
\multirow{2}{*}{Model} & HLS-Eval & \multicolumn{2}{c}{Bench4HLS} \\
\cmidrule(lr){2-2}\cmidrule(l){3-4}
 & p@1 & p@1 & p@5 \\
\midrule
GPT-5.6 Sol & $88.09\to \mathbf{98.94}$ & $77.88\to \mathbf{85.65}$ & $82.35\to \mathbf{91.76}$ \\
GPT-6.1 Sol & $96.81\to \mathbf{99.15}$ & $78.71\to \mathbf{84.59}$ & $83.53\to \mathbf{91.18}$ \\
Claude Opus 5.5 & $98.94\to \mathbf{100.00}$ & $77.41\to \mathbf{92.71}$ & $83.53\to \mathbf{93.53}$ \\
Claude Fable 5.1 & $99.79\to \mathbf{100.00}$ & $76.94\to \mathbf{92.00}$ & $85.88\to \mathbf{92.94}$ \\
\bottomrule
\end{tabular}
\end{table}

We further evaluate MementoHLS with four proprietary LLMs, as shown in Table~\ref{tab:closed_models}. On Bench4HLS, MementoHLS improves joint pass@1 and pass@5 by 5.88–15.29 and 7.06–10.00 percentage points, respectively. On HLS-Eval, joint pass@1 also improves, while pass@5 remains at 100\% for all models and is omitted for brevity. These results demonstrate that MementoHLS remains effective with strong backbones, complementing model capability with tool-grounded repair.

\renewcommand{\arraystretch}{1.08}
\setlength{\extrarowheight}{0.3pt}
\setlength{\tabcolsep}{2.2pt}


\newlength{\ApproachColWidth}
\newlength{\MetricColWidth}
\newlength{\MetricGroupWidth}

\setlength{\ApproachColWidth}{2.1cm}
\setlength{\MetricColWidth}{0.55cm}
\setlength{\MetricGroupWidth}{1.10cm}

\newcolumntype{C}[1]{>{\centering\arraybackslash}p{#1}}

\begin{table*}[t]
\centering
\caption{Component ablation of MementoHLS with the Llama-3-8B backbone on HLS-Eval and Bench4HLS.}
\label{tab:overall-results}

\scriptsize

\resizebox{\textwidth}{!}{%
\begin{tabular}{
@{}
>{\raggedright\arraybackslash}p{\ApproachColWidth}
*{20}{C{\MetricColWidth}}
@{}
}
\toprule


\multirow{3}{*}{\textbf{Variant}}
&
\multicolumn{10}{c}{\textbf{HLS-Eval \cite{AbiKaram2025HLSEval}} (\%)}
&
\multicolumn{10}{c}{\textbf{Bench4HLS \cite{Khan2026Bench4HLS}} (\%)}
\\

\cmidrule(lr){2-11}%
\cmidrule(lr){12-21}


&
\multicolumn{2}{c}{\makebox[\MetricGroupWidth][c]{\textbf{Parse}}}
&
\multicolumn{2}{c}{\makebox[\MetricGroupWidth][c]{\textbf{Compile}}}
&
\multicolumn{2}{c}{\makebox[\MetricGroupWidth][c]{\textbf{TB}}}
&
\multicolumn{2}{c}{\makebox[\MetricGroupWidth][c]{\textbf{Synth}}}
&
\multicolumn{2}{c}{\makebox[\MetricGroupWidth][c]{\textbf{TB$\wedge$Synth}}}
&
\multicolumn{2}{c}{\makebox[\MetricGroupWidth][c]{\textbf{Parse}}}
&
\multicolumn{2}{c}{\makebox[\MetricGroupWidth][c]{\textbf{Compile}}}
&
\multicolumn{2}{c}{\makebox[\MetricGroupWidth][c]{\textbf{TB}}}
&
\multicolumn{2}{c}{\makebox[\MetricGroupWidth][c]{\textbf{Synth}}}
&
\multicolumn{2}{c}{\makebox[\MetricGroupWidth][c]{\textbf{TB$\wedge$Synth}}}
\\

\cmidrule(lr){2-3}%
\cmidrule(lr){4-5}%
\cmidrule(lr){6-7}%
\cmidrule(lr){8-9}%
\cmidrule(lr){10-11}%
\cmidrule(lr){12-13}%
\cmidrule(lr){14-15}%
\cmidrule(lr){16-17}%
\cmidrule(lr){18-19}%
\cmidrule(lr){20-21}


&
\textbf{p@1} & \textbf{p@5}
&
\textbf{p@1} & \textbf{p@5}
&
\textbf{p@1} & \textbf{p@5}
&
\textbf{p@1} & \textbf{p@5}
&
\textbf{p@1} & \textbf{p@5}
&
\textbf{p@1} & \textbf{p@5}
&
\textbf{p@1} & \textbf{p@5}
&
\textbf{p@1} & \textbf{p@5}
&
\textbf{p@1} & \textbf{p@5}
&
\textbf{p@1} & \textbf{p@5}
\\

\midrule


Zero-shot$^{*}$
& 99.8 & 100.0
& 64.3 & 80.9
& 45.3 & 59.6
& 59.1 & 76.6
& 41.7 & 57.4
& 99.9 & 100.0
& 58.4 & 95.3
& 15.8 & 32.9
& 58.9 & 95.3
& 15.5 & 32.9
\\

Feedback
& 100.0 & 100.0
& 72.1 & 86.2
& 54.9 & 63.8
& 69.1 & 85.1
& 53.6 & 63.8
& 100.0 & 100.0
& 60.0 & 95.3
& 21.4 & 40.0
& 60.6 & 95.9
& 21.4 & 40.0
\\

RFL
& 100.0 & 100.0
& 78.5 & 90.4
& 55.7 & 67.0
& 76.0 & 90.4
& 54.5 & 67.0
& 100.0 & 100.0
& 66.1 & 95.9
& 21.8 & 40.6
& 66.8 & 96.5
& 21.5 & 40.6
\\

ERCL
& 100.0 & 100.0
& 78.3 & 92.6
& 55.7 & 68.1
& 77.2 & 92.6
& 55.7 & 68.1
& 100.0 & 100.0
& 73.3 & 100.0
& 26.8 & 46.5
& 73.5 & 98.8
& 26.7 & 46.5
\\

\textbf{MementoHLS}
& \textbf{100.0} & \textbf{100.0}
& \textbf{81.5}  & \textbf{91.5}
& \textbf{59.4}  & \textbf{69.1}
& \textbf{80.2}  & \textbf{92.6}
& \textbf{58.9}  & \textbf{68.1}
& \textbf{100.0} & \textbf{100.0}
& \textbf{79.8}  & \textbf{99.4}
& \textbf{32.5}  & \textbf{50.6}
& \textbf{79.2}  & \textbf{98.2}
& \textbf{32.4}  & \textbf{50.6}
\\

\bottomrule
\end{tabular}%
}

\vspace{2pt}
\begin{minipage}{\textwidth}
\scriptsize
$^{*}$Zero-shot denotes direct Round-0 generation without iterative feedback or repair.
\end{minipage}

\end{table*}

\subsection{ERCL and RFL Ablation}

To isolate the contribution of each component, we fix Llama-3-8B as the
backbone for all RQ2 experiments. All configurations share the same Round-0
candidates and repair budget. We compare four iterative settings: Feedback
without memory, RFL only, ERCL only, and the complete MementoHLS with both
components enabled. These configurations form a $2\times2$ factorial design
over ERCL and RFL, while Zero-shot serves only as the backbone reference.


As shown in Table~\ref{tab:overall-results}, ERCL and RFL improve
different aspects of the repair process. RFL uses tool outcomes to
preserve validated progress and suppress ineffective or regressive
repair routes. ERCL applies evidence-gated repair contracts to the
current failure. The full framework achieves the highest joint
pass@1 on both benchmarks. Adding RFL to ERCL-only repair raises
joint pass@1 by 3.2 percentage points on HLS-Eval and 5.7 points
on Bench4HLS. Adding RFL to feedback-only repair yields gains of
0.9 and 0.1 points, respectively. On HLS-Eval, ERCL-only repair
and the full framework both achieve 68.1\% joint pass@5. Each component provides consistent benefits independently, while their combination achieves the strongest TB$\land$Synth performance on both benchmarks.

These results demonstrate that ERCL and RFL play complementary roles in
MementoHLS. ERCL determines which reusable repair knowledge is applicable to
the current failure, whereas RFL evaluates previous repair outcomes based on
actual tool feedback and constrains subsequent repair decisions. By jointly
leveraging cross-task repair knowledge and task-local trajectory evidence,
MementoHLS achieves more reliable multi-round HLS repair.

\begin{figure}[t]
    \centering
    \includegraphics[width=1\linewidth]{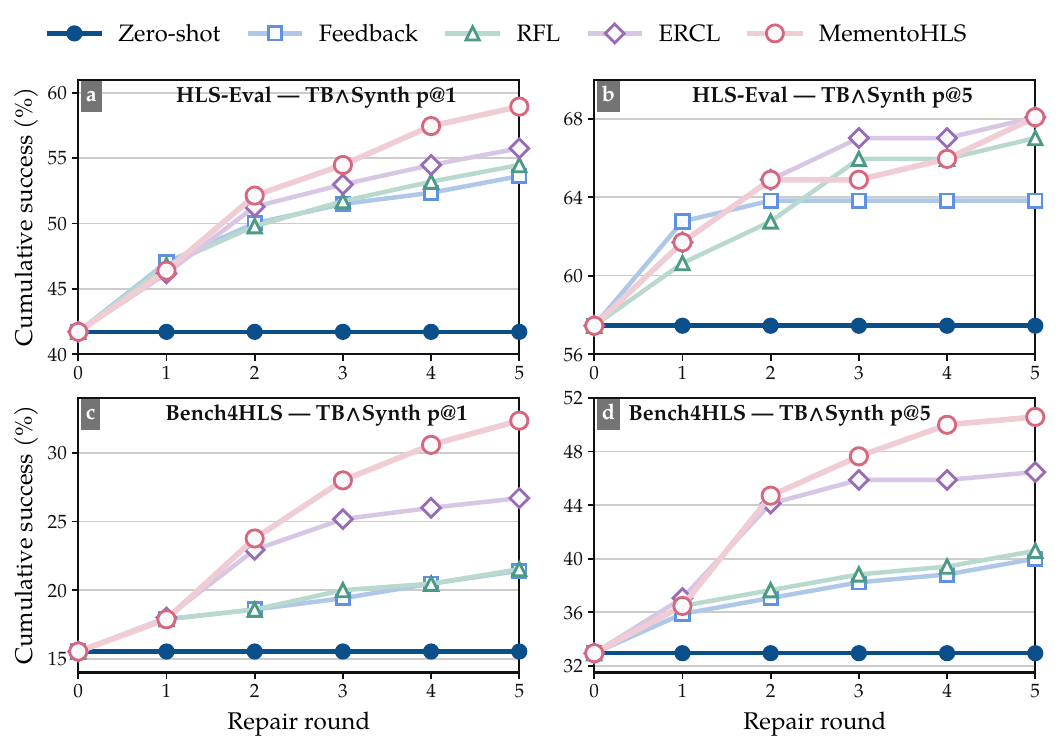}
    \vspace{-0.6cm}
    \caption{Cumulative TB$\land$Synth pass@1/pass@5 by repair round (Llama-3-8B).}
    \label{fig:RQ3_fig1}
    \vspace{-0.4cm}
\end{figure}

\begin{table}[t]
\centering
\caption{Latency comparison with manually optimized benchmark designs (Llama-3-8B, \%)}
\label{tab:optimization_outcomes}

\begin{threeparttable}

\footnotesize
\setlength{\tabcolsep}{0.5pt}
\renewcommand{\arraystretch}{1.05}

\sisetup{
    detect-weight = true,
    detect-family = true,
    table-number-alignment = center
}

\begin{tabular*}{\columnwidth}{
    @{\extracolsep{\fill}}
    >{\raggedright\arraybackslash}p{0.25\columnwidth}
    S[table-format=2.2, table-column-width=0.09\columnwidth]
    S[table-format=2.2, table-column-width=0.09\columnwidth]
    S[table-format=2.2, table-column-width=0.09\columnwidth]
    S[table-format=2.2, table-column-width=0.09\columnwidth]
    S[table-format=1.2e1, table-column-width=0.145\columnwidth]
    @{}
}
\toprule

\textbf{Setting}
& \multicolumn{1}{c}{\makecell{\textbf{Not}$^*$\\\textbf{Slower} $\uparrow$}}
& \multicolumn{1}{c}{\makecell{\textbf{Faster}\\$\uparrow$}}
& \multicolumn{1}{c}{\makecell{\textbf{Slower}\\$\downarrow$}}
& \multicolumn{1}{c}{\makecell{\textbf{Avg.}\\\textbf{Red.} $\uparrow$}}
& \multicolumn{1}{c}{\makecell{\textbf{Avg.}\\\textbf{Inc.} $\downarrow$}}
\\
\midrule

\multicolumn{6}{@{}l}{
    \textit{\textbf{HLS-Eval}}~\cite{AbiKaram2025HLSEval}
} \\[-0.3ex]

\hspace{0.5em}Zero-shot
& 66.56
& 46.98
& 33.44
& 68.26
& {1.01e3}
\\

\hspace{0.5em}\textbf{MementoHLS}
& {\bfseries 75.68}
& {\bfseries 47.30}
& {\bfseries 24.32}
& {\bfseries 75.02}
& {\bfseries 7.23e2}
\\

\addlinespace[1pt]
\midrule
\addlinespace[0.3pt]

\multicolumn{6}{@{}l}{
    \textit{\textbf{Bench4HLS}}~\cite{Khan2026Bench4HLS}
} \\[-0.3ex]

\hspace{0.5em}Zero-shot
& 86.99
& 2.08
& 13.01
& 91.44
& {2.49e8}
\\

\hspace{0.5em}\textbf{MementoHLS}
& {\bfseries 87.18}
& {\bfseries 3.21}
& {\bfseries 12.82}
& {\bfseries 94.87}
& {\bfseries 4.50e3}
\\

\bottomrule
\end{tabular*}

\vspace{1pt}

\begin{tablenotes}[flushleft]
\scriptsize
\item[]
* Not Slower denotes designs with latency less than or equal to the
reference. Avg. Red./Inc. denotes the average latency reduction/increase
among designs faster/slower than the reference, respectively.
\end{tablenotes}

\end{threeparttable}
\end{table}

\begin{figure}[t]
    \centering
    \includegraphics[width=1\linewidth]{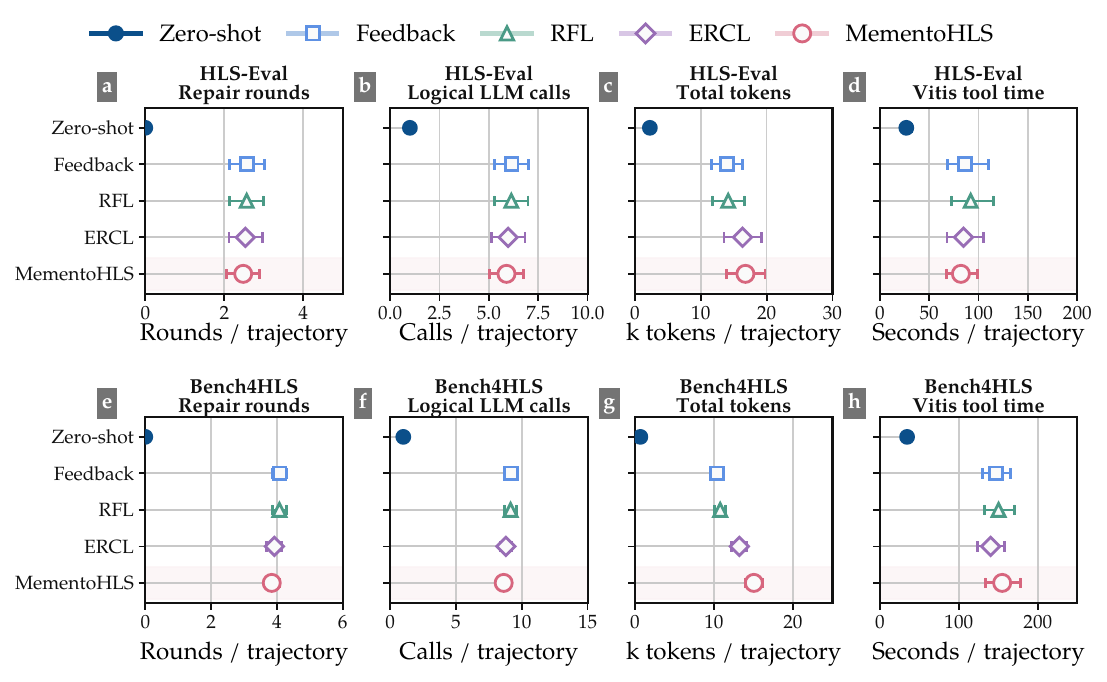}
    \vspace{-0.6cm}
    \caption{Per-trajectory cost by configuration (Llama-3-8B).}
    \label{fig:RQ3_fig3}
    \vspace{-0.4cm}
\end{figure}

\subsection{Multi-Round Repair: Gains and Cost}

\textit{Round-wise Repair Gains.}
Fig.~\ref{fig:RQ3_fig1} shows that most gains are obtained in the early repair rounds. On Bench4HLS, the first three rounds account for 74.1\% and 83.3\% of the final pass@1 and pass@5 improvements, respectively; the corresponding ratios on HLS-Eval are 74.1\% and 70.0\%. 
Most of the total improvement is achieved within the first three rounds, while Round 5 still yields additional gains on both benchmarks.
Moreover, the complete MementoHLS continues to make progress in later rounds, whereas the single-component variants tend to plateau earlier. These results indicate that most failures can be repaired within the first few rounds, while the complete system retains the ability to recover additional difficult cases in later rounds.


\textit{Repair Efficiency.}
Fig.~\ref{fig:RQ3_fig3} reports per-trajectory costs for Llama-3-8B.
Each repair trajectory runs for at most five rounds and stops once a candidate passes both TB and Synth.
MementoHLS averages 3.84 repair rounds and 8.62 logical LLM calls per trajectory on Bench4HLS, and 2.48 rounds and 5.89 calls on HLS-Eval, the lowest averages among the iterative configurations.
Its joint pass@1 and pass@5 gains over Zero-shot range from 10.7 to 17.7 percentage points across the two benchmarks.
Compared with Feedback, MementoHLS makes fewer logical LLM calls per trajectory and raises joint pass@1 from 21.4\% to 32.4\% on Bench4HLS and from 53.6\% to 58.9\% on HLS-Eval.


\textit{Latency of Generated Designs.}
We evaluate latency only for generated designs that pass synth on the same candidate, with latency measured through HLS RTL simulation.
As shown in Table~\ref{tab:optimization_outcomes}, the fraction of these valid designs that are not slower than the benchmark reference increases from 66.56\% to 75.68\% on HLS-Eval and from 86.99\% to 87.18\% on Bench4HLS.
These results indicate that MementoHLS improves end-to-end correctness while maintaining a high proportion of latency-competitive designs, providing valid starting points for subsequent latency, resource, and PPA optimization.




\section{Conclusion and Future Work}

We introduced MementoHLS, an inference-time repair framework for
reliable LLM-driven HLS generation. It combines executable repair
contracts from ERCL with trajectory evidence from RFL in an
evidence-driven closed loop around a fixed LLM. With up to five
repair rounds, MementoHLS improves same-candidate TB$\wedge$Synth
pass@1/pass@5 over Zero-shot by an average of 16.1/12.1 percentage
points on HLS-Eval and 15.1/10.4 percentage points on Bench4HLS
across Llama-3-8B and Qwen2.5-Coder-7B/32B. On four proprietary LLMs,
the average Bench4HLS gains are 11.0/8.5 percentage points.
These results demonstrate benefits across model families and
parameter scales, while the Llama-3-8B ablations support the
complementary roles of ERCL and RFL. Future work will explore
additional HLS toolchains, unseen algorithm families, and stronger
correctness and QoR-aware repair under controlled resource budgets.


\clearpage
\bibliographystyle{IEEEtran}
\bibliography{reference.bib}

@article{Chen2024Allo,
  author  = {Chen, Hongzheng and Zhang, Niansong and Xiang, Shaojie and others},
  title   = {{Allo}: A Programming Model for Composable Accelerator Design},
  journal = {Proceedings of the ACM on Programming Languages},
  volume  = {8},
  number  = {{PLDI}},
  pages   = {593--620},
  year    = {2024}
}

@inproceedings{europar26-two,
  title={Two-Stage Hierarchy-Aware Learning with Gradient Conflict Mitigation for HLS Latency and Resource Prediction},
  author={Ke, Zhiwei and Liu, Yiming and Zuo, Fengrui and Lou, Wenqi and Wang, Chao and Zhou, Xuehai},
  booktitle={European Conference on Parallel Processing},
  pages={394--408},
  year={2026},
  organization={Springer}
}

@inproceedings{Ye2024HIDA,
  author    = {Ye, Hanchen and Jun, Hyegang and Chen, Deming},
  title     = {{HIDA}: A Hierarchical Dataflow Compiler for
               High-Level Synthesis},
  booktitle = {Proceedings of the ACM International Conference on Architectural Support for Programming Languages and Operating Systems},
  pages     = {215--230},
  year      = {2024}
}

@inproceedings{AbiKaram2025HLSEval,
  author    = {Abi-Karam, Stefan and Hao, Cong},
  title     = {{HLS-Eval}: A Benchmark and Framework for Evaluating
               {LLM}s on High-Level Synthesis Design Tasks},
  booktitle = {Proceedings of the IEEE International Conference on LLM-Aided Design},
  year      = {2025},
  pages     = {219--226}
}

@inproceedings{Xing2026ChipMind,
  author    = {Xing, Changwen and Wong, SamZaak and Wan, Xinlai and others},
  title     = {{ChipMind}: Retrieval-Augmented Reasoning for
               Long-Context Circuit Design Specifications},
  booktitle = {Proceedings of the AAAI Conference on Artificial Intelligence},
  volume    = {40},
  number    = {2},
  pages     = {1337--1345},
  year      = {2026}
}

@inproceedings{Shinn2023Reflexion,
  author    = {Shinn, Noah and Cassano, Federico and Gopinath, Ashwin and others},
  title     = {{Reflexion}: Language Agents with Verbal Reinforcement Learning},
  booktitle = {Advances in Neural Information Processing Systems},
  volume    = {36},
  pages     = {8634--8652},
  year      = {2023}
}

@inproceedings{Khan2026Bench4HLS,
  author    = {Khan, M. Zafir Sadik and Azar, Kimia Zamiri and Kamali, Hadi},
  title     = {{Bench4HLS}: End-to-End Evaluation of {LLM}s in
               High-Level Synthesis Code Generation},
  booktitle = {Proceedings of the Design, Automation \& Test in Europe Conference},
  year      = {2026},
  pages     = {1--7}
}

@inproceedings{Khan2025SAGEHLS,
  author    = {Khan, M. Zafir Sadik and Mashnoor, Nowfel and Akyash, Mohammad and others},
  title     = {{SAGE-HLS}: Syntax-Aware {AST}-Guided {LLM} for
               High-Level Synthesis Code Generation},
  booktitle = {Proceedings of the IEEE International Conference on Computer Design},
  year      = {2025},
  pages     = {574--581}
}

@article{Prakriya2025LIFT,
  author  = {Prakriya, Neha and Ding, Zijian and Sun, Yizhou and others},
  title   = {{LIFT}: {LLM}-Based Pragma Insertion for {HLS} via
             {GNN} Supervised Fine-Tuning},
  journal = {arXiv:2504.21187},
  year    = {2025}
}

@article{Zou2026HLSSeek,
  author  = {Zou, Qingyun and Yu, Feng and Tan, Hongshi and others},
  title   = {{HLS-Seek}: {QoR}-Aware Code Generation for High-Level
             Synthesis via Proxy Comparative Reward Reinforcement Learning},
  journal = {arXiv:2605.13536},
  year    = {2026}
}

@inproceedings{Xiong2024HLSPilot,
  author    = {Xiong, Chenwei and Liu, Cheng and Li, Huawei and others},
  title     = {{HLSPilot}: {LLM}-Based High-Level Synthesis},
  booktitle = {Proceedings of the IEEE/ACM International Conference on Computer-Aided Design},
  year      = {2024},
  pages     = {1--9}
}

@inproceedings{Xu2024RALAD,
  author    = {Xu, Haocheng and Hu, Haotian and Huang, Sitao},
  title     = {Optimizing High-Level Synthesis Designs with
               Retrieval-Augmented Large Language Models},
  booktitle = {Proceedings of the IEEE LLM Aided Design Workshop},
  year      = {2024},
  pages     = {1--5}
}

@article{Collini2025C2HLSC,
  author  = {Collini, Luca and Garg, Siddharth and Karri, Ramesh},
  title   = {{C2HLSC}: Leveraging Large Language Models to Bridge
             the Software-to-Hardware Design Gap},
  journal = {ACM Trans. Des. Autom. Electron. Syst.},
  volume  = {30},
  number  = {6},
  year    = {2025},
  pages   = {96:1--96:24}
}

@article{Xu2026HLSRewriter,
  author  = {Xu, Kangwei and Zhang, Grace Li and Yin, Xunzhao and others},
  title   = {{HLSRewriter}: Efficient Refactoring and Optimization of
             {C/C++} Code with {LLM}s for High-Level Synthesis},
  journal = {ACM Trans. Des. Autom. Electron. Syst.},
  volume  = {31},
  number  = {4},
  year    = {2026},
  pages   = {61:1--61:21}
}

@inproceedings{tsai2024rtlfixer,
  title={{RTLFixer: Automatically Fixing RTL Syntax Errors with Large Language Model}},
  author={Tsai, Yun-Da and Liu, Mingjie and Ren, Haoxing},
  booktitle={Proceedings of the 61st ACM/IEEE Design Automation Conference},
  pages={1--6},
  year={2024}
}

@article{wu2024chateda,
  title={{ChatEDA: A Large Language Model Powered Autonomous Agent for EDA}},
  author={Wu, Haoyuan and He, Zhuolun and Zhang, Xinyun and Yao, Xufeng and Zheng, Su and Zheng, Haisheng and Yu, Bei},
  journal={IEEE Transactions on Computer-Aided Design of Integrated Circuits and Systems},
  year={2024},
  volume={43},
  number={10},
  pages={3184--3197},
  publisher={IEEE}
}

@inproceedings{Xu2024MEIC,
  author    = {Ke Xu and
               Jialin Sun and
               Yuchen Hu and
               Xinwei Fang and
               Weiwei Shan and
               Xi Wang and
               Zhe Jiang},
  title     = {{MEIC:} Re-thinking {RTL} Debug Automation using LLMs},
  booktitle = {Proceedings of the 43rd {IEEE/ACM} International Conference on Computer-Aided
               Design},
  pages     = {1--9},
  year      = {2024}
}

@inproceedings{Li2026ChatHLS,
  author    = {Runkai Li and Jia Xiong and Xiuyuan He and Jieru Zhao and Jiaqi Lv and Haowen Fang and Lei Qi and Xi Wang},
  title     = {{ChatHLS}: Towards Systematic Design Automation and
               Optimization for High-Level Synthesis},
  booktitle = {Proceedings of the 64th Annual Meeting of the Association for Computational Linguistics},
  year      = {2026},
  pages     = {20996--21015}
}

@inproceedings{Zou2026AgRefactor,
author = {Zou, Yang and Ding, Zijian and Wang, Chi and Sun, Yizhou and Cong, Jason},
title = {AgRefactor: Refactoring for HLS Compatibility with a Self-Evolving Agentic Workflow},
year = {2026},
booktitle = {Proceedings of the 2026 ACM/SIGDA International Symposium on Field Programmable Gate Arrays},
pages = {183},
numpages = {1}
}

@misc{mementohls,
  author       = {{MementoHLS}},
  title        = {{MementoHLS} Replication Package},
  howpublished = {\url{https://github.com/ZD-zhu/MementoHLS}},
  year         = {2026}
}

@inproceedings{Wang2025AgentWorkflowMemory,
  author    = {Wang, Zora Zhiruo and Mao, Jiayuan and Fried, Daniel and others},
  title     = {Agent Workflow Memory},
  booktitle = {Proceedings of the International Conference on Machine Learning},
  volume    = {267},
  pages     = {63897--63911},
  year      = {2025}
}

@inproceedings{gai2025exploring,
  author    = {Jiahao Gai and Hao Chen and Zhican Wang and others},
  title     = {Exploring Code Language Models for Automated {HLS}-Based Hardware Generation: Benchmark, Infrastructure and Analysis},
  booktitle = {Proceedings of the Asia and South Pacific Design Automation Conference},
  year      = {2025},
  pages     = {988--994}
}

@article{zhang2026contracthil,
  author  = {Jingbo Zhang and Haoxiang Sun and Wenbo Wang and others},
  title   = {{ContractHIL-HLS}: Contract-Aligned Multi-Agent Workflow with Hardware-in-the-Loop Feedback for {HLS} Design},
  journal = {arXiv:2607.25283},
  year    = {2026}
}

@article{Blocklove2025AutoChip,
  author  = {Blocklove, Jason and Thakur, Shailja and Tan, Benjamin and
             Pearce, Hammond and Garg, Siddharth and Karri, Ramesh},
  title   = {Automatically Improving {LLM}-based {Verilog} Generation
             using {EDA} Tool Feedback},
  journal = {ACM Trans. Des. Autom. Electron. Syst.},
  volume  = {30},
  number  = {6},
  year    = {2025},
  pages   = {100:1--100:26}
}

@inproceedings{Ho2025VerilogCoder,
  author    = {Ho, Chia-Tung and Ren, Haoxing and Khailany, Brucek},
  title     = {{VerilogCoder}: Autonomous {Verilog} Coding Agents with
               Graph-based Planning and Abstract Syntax Tree ({AST})-based
               Waveform Tracing Tool},
  booktitle = {Proceedings of the {AAAI} Conference on Artificial Intelligence},
  volume    = {39},
  number    = {1},
  year      = {2025},
  pages     = {300--307}
}

@inproceedings{Xu2025HLSTester,
  author    = {Xu, Kangwei and Li, Bing and Zhang, Grace Li and Schlichtmann, Ulf},
  title     = {{HLSTester}: Efficient Testing of Behavioral Discrepancies
               with {LLM}s for High-Level Synthesis},
  booktitle = {Proceedings of the 44th {IEEE/ACM} International Conference
               on Computer-Aided Design},
  year      = {2025},
  doi       = {10.1109/ICCAD66269.2025.11240891}
}

@inproceedings{Zhao2025MAGE,
  author    = {Zhao, Yujie and Zhang, Hejia and Huang, Hanxian and
               Yu, Zhongming and Zhao, Jishen},
  title     = {{MAGE}: A Multi-Agent Engine for Automated {RTL} Code Generation},
  booktitle = {Proceedings of the 62nd {ACM/IEEE} Design Automation Conference},
  year      = {2025},
  pages     = {1--7},
  doi       = {10.1109/DAC63849.2025.11133191}
}

@article{Chen2026ReflexHLS,
  author  = {Chen, Delang and Luo, Jizhou and Liu, Zhongli and
             Li, Xingji and Yang, Jianqun and Liu, Jiaxi},
  title   = {Constraint-Aware {HLS} Optimization via {LLM}-Based
             Multi-Role Agent with Iterative Reflexion},
  journal = {ACM Trans. Des. Autom. Electron. Syst.},
  year    = {2026},
  doi     = {10.1145/3849869}
}

\appendices
\begingroup
\makeatletter
\renewcommand{\section}{%
  \@startsection{section}{1}{\z@}%
    {-1.25ex \@plus -.4ex \@minus -.2ex}%
    {.6ex \@plus .2ex}%
    {\normalfont\large\bfseries\raggedright}}
\def\thesectiondis{\thesection}
\def\@seccntformat#1{\csname the#1\endcsname\quad}
\renewcommand{\paragraph}[1]{%
  \par\addvspace{1ex plus .2ex minus .15ex}%
  \noindent\textbf{#1}\enspace\ignorespaces
}
\def\fnum@table{Table~\thetable}
\def\fnum@figure{Figure~\thefigure}
\def\@makecaption#1#2{%
  \vskip\abovecaptionskip
  {\normalfont\footnotesize\bfseries\raggedright #1: #2\par}%
  \vskip\belowcaptionskip
}
\makeatother

\clearpage
\begingroup
\setlength{\extrarowheight}{0pt}
\renewcommand{\arraystretch}{1.05}
\setcounter{table}{0}
\setcounter{figure}{0}
\renewcommand{\thetable}{\Alph{section}.\arabic{table}}
\renewcommand{\thefigure}{\Alph{section}.\arabic{figure}}
\renewcommand{\theHtable}{appendix.\thetable}
\renewcommand{\theHfigure}{appendix.\thefigure}

\section{Quality of Results on Nontrivial Kernels}
\label{app:kernel-qor}

Selected Llama-3-8B case studies show resource savings on a
1024-term fixed-point dot product and a 1024-bit, 256-way selector.
A multi-cycle image stencil shows lower latency with a small LUT
trade-off. M denotes MementoHLS; FT denotes the existing LoRA model.

\paragraph{Measurement scope.}
All reports use Vitis HLS 2024.2, \texttt{xczu9eg-}\allowbreak
\texttt{ffvb1156-2-e}, and a 5\,ns target clock.
Reference is the benchmark kernel, not the testbench driver, and is
synthesized once. Cycles are top-level worst-case HLS estimates;
$T_{\rm est}$ is the estimated clock period in ns, not end-to-end time.
Resource counts and latency always refer to the same implementation.

\begin{table}[!htbp]
\centering
\footnotesize
\setlength{\tabcolsep}{4pt}
\renewcommand{\arraystretch}{1.10}
\caption{HLS-Eval: lower resource use for a 1024-term arithmetic reduction without increased cycle latency.}
\label{tab:app-hls-kernel-qor}

\begin{tabular*}{\linewidth}{@{\extracolsep{\fill}}lrrrrr@{}}
\toprule
Method & Cycles & LUT & FF & DSP & $T_{\rm est}\,(\mathrm{ns})$ \\
\midrule
\multicolumn{6}{@{}l}{\textbf{1024-term fixed-point dot product}} \\
Reference & 37 & 1,809 & 2,675 & 64 & 3.420 \\
Base & 1,028 & 299 & 175 & 2 & 3.420 \\
\textbf{Base + M} & 1,028 & 156 & 145 & 2 & 3.420 \\
FT & 516 & 2,468 & 807 & 4 & 3.613 \\
\textbf{FT + M} & 516 & 2,468 & 807 & 4 & 3.613 \\
\bottomrule
\end{tabular*}

\par\smallskip
\begin{minipage}{\linewidth}
\footnotesize\raggedright
All numeric rows pass TB and synthesis. Reference is deterministic.
$T_{\rm est}$ is in ns. LUT, FF and DSP are absolute counts;
BRAM and URAM are zero. Bold identifies M configurations,
not statistical significance.
\end{minipage}
\end{table}

\paragraph{Arithmetic reduction.}
The dot product retains all 1024 products and their accumulation,
using the benchmark's 32-bit fixed-point parameter/result type and
16-bit fixed-point feature type. Base + M uses 156 rather than 299 LUT
(47.8\% fewer) and 145 rather than 175 FF (17.1\% fewer), while retaining
1,028 cycles, two DSP blocks, and a 3.420\,ns estimated clock period.
Its selected candidate is from repair Round~3. The FT path remains
unchanged at 516 cycles and four DSP blocks. Thus, the benefit on this
case is a smaller Base-path footprint, not fewer execution cycles.
The reference uses explicit 32-way array partitioning/parallelism
and 64 DSP blocks; these different architectures must not be interpreted
as evidence that M is faster than the reference.

\newpage

\begin{table}[!htbp]
\centering
\footnotesize
\setlength{\tabcolsep}{4pt}
\renewcommand{\arraystretch}{1.10}
\caption{Bench4HLS: a multi-cycle image stencil with lower latency and a small LUT trade-off.}
\label{tab:app-bench-kernel-qor}

\begin{tabular*}{\linewidth}{@{\extracolsep{\fill}}lrrrrr@{}}
\toprule
Method & Cycles & LUT & FF & DSP & $T_{\rm est}\,(\mathrm{ns})$ \\
\midrule
\multicolumn{6}{@{}l}{\textbf{$5{\times}5$ image-window sum (\texttt{Prob156})}} \\
Reference & 77 & 5,545 & 5,298 & 0 & 3.529 \\
Base & 939 & 2,484 & 689 & 0 & 3.633 \\
\textbf{Base + M} & 872 & 2,575 & 603 & 0 & 3.242 \\
FT & 304 & 3,643 & 3,053 & 0 & 3.607 \\
\textbf{FT + M} & 304 & 3,643 & 3,053 & 0 & 3.607 \\
\bottomrule
\end{tabular*}

\par\smallskip
\begin{minipage}{\linewidth}
\footnotesize\raggedright
All rows pass TB and synthesis. $T_{\rm est}$ is in ns.
BRAM and URAM are zero. The Base-path LUT increase is
91 (3.7\%); this is not an all-metric non-regression example.
\end{minipage}
\end{table}

\paragraph{Multi-cycle image stencil.}
The \texttt{Prob156} kernel computes all 64 output pixels of an
$8{\times}8$ image, each with a bounded $5{\times}5$ neighborhood
and zero-padding boundary semantics. Base + M reduces cycle latency
from 939 to 872 (7.1\%) and FF from 689 to 603 (12.5\%);
$T_{\rm est}$ decreases from 3.633 to 3.242\,ns.
LUT increases from 2,484 to 2,575: 91 LUT, or 3.7\%.
This is explicitly a latency improvement with a small LUT trade-off,
not an all-metric non-regression example. The FT path is unchanged
at 304 cycles. The selected Base + M implementation is from Round~2.

\begin{table}[!htbp]
\centering
\footnotesize
\setlength{\tabcolsep}{4pt}
\renewcommand{\arraystretch}{1.10}
\caption{Bench4HLS: resource savings in a wide selection datapath.}
\label{tab:app-bench-wide-qor}

\begin{tabular*}{\linewidth}{@{\extracolsep{\fill}}lrrrrr@{}}
\toprule
Method & Cycles & LUT & FF & DSP & $T_{\rm est}\,(\mathrm{ns})$ \\
\midrule
\multicolumn{6}{@{}l}{\textbf{1024-bit, 256-to-1 four-bit selector (\texttt{Prob021})}} \\
Reference & 0 & 273 & 0 & 0 & 1.362 \\
Base & -- & -- & -- & -- & -- \\
\textbf{Base + M} & 0 & 2,171 & 0 & 0 & 1.887 \\
FT & 0 & 2,171 & 0 & 0 & 1.887 \\
\textbf{FT + M} & 0 & 273 & 0 & 0 & 1.362 \\
\bottomrule
\end{tabular*}

\par\smallskip
\begin{minipage}{\linewidth}
\footnotesize\raggedright
All numeric rows pass TB and synthesis. $T_{\rm est}$ is in ns.
A dash denotes no eligible candidate, not zero.
No Base-path PPA comparison is claimed.
BRAM and URAM are zero.
Zero cycles do not mean zero physical delay.
\end{minipage}
\end{table}

\paragraph{Wide-datapath selection.}
\texttt{Prob021} selects one of 256 four-bit fields from a 1024-bit input.
FT + M reduces LUT from 2,171 to 273 (87.4\%) and $T_{\rm est}$
from 1.887 to 1.362\,ns (27.8\%), matching the reference resource
and clock estimates.
FF and DSP remain zero, and the reported cycle latency stays zero.
The selected FT implementation uses a wide shift followed by masking;
FT + M, selected at Round~1, uses a variable four-bit slice.
Both preserve the selector-to-bit-field relation. The result is
a wide combinational datapath area/timing benefit, not multi-cycle
acceleration. Interface pragmas can differ, so the counts describe
complete HLS implementations rather than an isolated fixed-interface
datapath rewrite.

\paragraph{Interpretation.}
The positive evidence is confined to these best-observed implementations:
resource savings with preserved cycle latency, plus the explicitly
qualified latency--LUT trade-off for the image stencil.

\clearpage

\twocolumn[{%
\begin{minipage}{\textwidth}

\section{Additional Multi-Round Repair Results}
\label{app:marginal-repair-gains}

Figure~\ref{fig:app-marginal-gains} provides a per-round view of the
existing repair-budget evaluation. For each $k\in\{1,5\}$, the marginal
gain at round $t$ is the cumulative TB\&Synth pass@$k$ at $t$ minus
that at $t-1$, expressed in percentage points. The five methods use the
same evaluation setting as the main-text repair-budget experiment.
Zero-shot has no repair-stage gain. This view complements the main-text
cumulative and cost plots; it introduces neither an additional
experiment nor a new candidate-selection policy.

\bigskip
\centering
\includegraphics[width=0.98\textwidth]{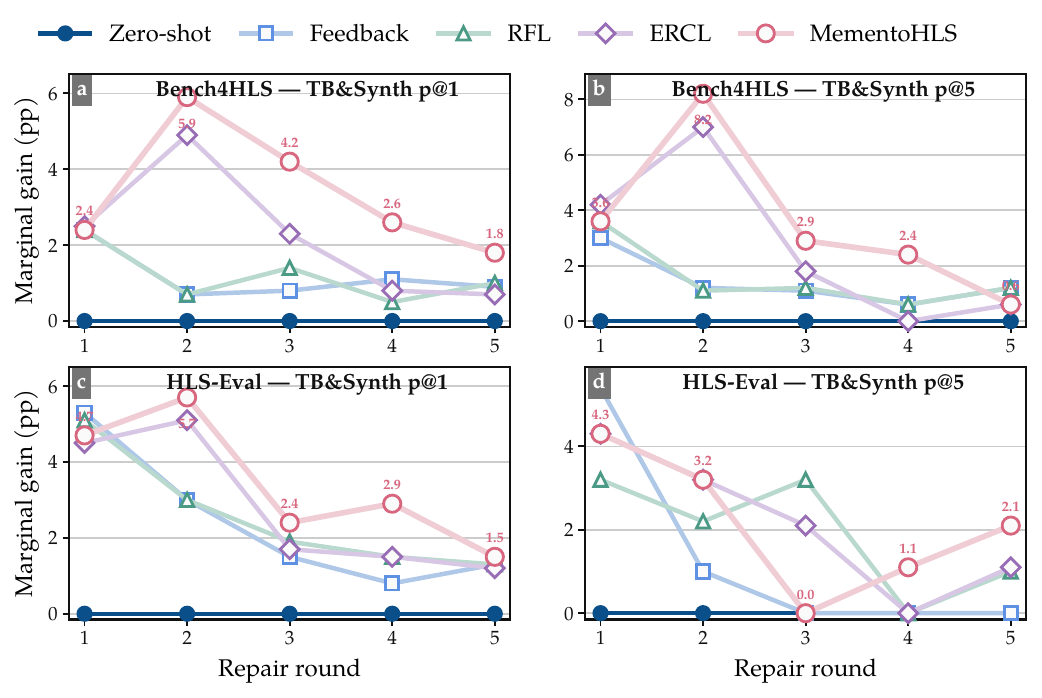}

\captionof{figure}{Per-round marginal TB\&Synth gains on Bench4HLS
and HLS-Eval. Each panel reports changes in pass@1 or pass@5 over
rounds 1--5. Values are percentage-point increments, not relative
percentage improvements.}

\Description{Four panels show per-round TB and synthesis pass-rate
gains for five methods on Bench4HLS and HLS-Eval at pass@1 and pass@5.}

\label{fig:app-marginal-gains}

\end{minipage}
}]

\endgroup

\endgroup

\end{document}